# Numerical Solution of the Schrödinger Equation for a Short-Range 1/r Singular Potential with any ℓ Angular Momentum

**Abdulla Jameel Sous [1], M. I. El-Kawni [2]**

1. Department of Mathematics, Faculty of Technology and Applied Sciences, Al-Quds Open University, Tulkarm, Palestine, email: asous@qou.edu
2. Department of Science, Faculty of Technology and Applied Sciences, Al-Quds Open University, Nablus, Palestine, email: mkawni@qou.edu

**Abstract:** *Recently, the Asymptotic Iteration Method (AIM) was used to calculate the energy spectrum for a short rang three parameter central potential which was introduced by H. Bahlouli and A. D. Alhaidari. The S-orbital wave solution of the Schrödinger equation was obtained for different parameters of the potential. In this work a non-zero angular momentum term were introduced to the problem and the energy eigenvalues were obtained for different potential parameters. Our results show very good agreements compared with other methods such as potential parameter spectrum method (PPSM) and the complex scaling method (CSM)*.

## INTRODUCTION

In quantum mechanics, The most basic problem is to solve the Schrödinger equation for the energy eigenvalues $E_n$ and the associated energy Eigen functions.

There are a number of important cases for which the stationary Schrödinger equation can be solved analytically. However, analytic solutions are possible only for a few simple quantum systems such as the hydrogen atom, the harmonic oscillator and others [1-2]. In most cases, many quantum systems can be treated only by approximation methods. Among such approximation methods include Pekeris approximation [3], Semi-classical method [4], and asymptotic iteration method [5-10].

Recently, the study of exponential-type potentials has attracted much attention from many authors [11-13]. These potentials include the Hulthén potential [14], the multi parameter exponential-type potentials [15], the Manning–Rosen potential [13] and the Eckart-type potential [16]. It should be mentioned that most contributions appearing in the literature are concerned with the s- wave case.

In this work we study the arbitrary $\ell$ - state solutions of the Schrödinger equation with a short rang three parameter central potential which was introduced by H. Bahlouli and A. D. Alhaidari [17-18]

$$V(r) = V_0 \frac{e^{-\lambda r} - \gamma}{e^{\lambda r} - 1} \quad (1)$$

Where $V_0$ is the potential strength, and the range parameter $\lambda$ is positive with an inverse length units. The dimensionless parameter $\gamma$ is in the open range $0 < \gamma < 1$. This potential is short-range with $1/r$ singularity at the origin. It is also interesting to note that, at short distance and with $0 < \gamma < 1$, there is a clear resemblance of this potential with $V_0 > 0$ to the attractive Coulomb potential [5] with non-zero angular momentum. The potential valley here is not due to the centrifugal force attributed to the angular momentum. Moreover, it does not have the long-range behavior of the Coulomb potential [17, 18]. In [18], Alhaidari argued that in contrast to the Coulomb potential the number of bound states for this potential is finite and that it could be used as a more appropriate model for the description of an electron interacting with an extended molecule whose electron cloud is congregated near the center of the molecule. The Authors of [17-19] found the "potential parameter spectrum" (PPS) for the hyperbolic single wave potential and for potential (1). The concept of a PPS was introduced for the first time in the solution of the wave equation in [20] where for a given energy the problem becomes exactly solvable for a discrete set (finite or infinite) of values of the potential parameters. If the map that associates the parameter spectrum with the energy is invertible, then in principle one could obtain the energy spectrum for a given choice of potential parameters [20].

In a previous article [19], we used the Asymptotic Iteration Method (AIM) to find the energy spectrum for the s-wave (zero angular momentum) Schrodinger Equation with potential in (1) which was introduced by Bahlouli and Alhaidari [17- 18]. In the present work, we apply the same technique in [5- 10] to the same potential in (1), and we will find the eigenvalues for the time-independent radial Schrödinger equation for any angular momentum $\ell$ .

The paper has the following structures. In Section 2, we briefly present an overview of the AIM which introduced to find the solutions for the second-order differential equation. In Sec. 3, change of variables and approximation scheme has been done which allows as transforming Schrödinger equation to another form in order to apply the method to solve the equation with the short rang three parameter central potential, In Sec. 3, our numerical calculation results have been presented for the eigenvalues.

## CONCEPT OF ASYMPTOTIC ITERATION METHOD

In this section we shall outline the general procedure of the AIM for determining the eigenvalue differential equation. The AIM has been proposed and used to solve the homogenous linear second-order differential equation of the form [5- 10],

$$f_n^{''}(x) = \lambda_0(x) f_n^{'}(x) + s_0(x) f_n(x) \tag{2}$$

Where $f(x)$ is a function of $x$ , $f^{/}(x), f^{//}(x)$ are the first and second derivatives with respect to $x$, $\lambda_0(x)$ , $s_0(x)$ are arbitrary functions in $C_\infty(a,b)$ and $\lambda_0(x) \neq 0$. To find a general solution, equation (2) can be iterated up to the up $(k+1)\,th$ and $(k+2)\,th$ derivatives, where $k = 1,2,3,...$ is the iteration number. Then one obtains

$$\begin{aligned} f_n^{(k+1)}(x) &= \lambda_{k-1}(x) f_n^{'}(x) + s_{k-1}(x) f_n(x) \\ f_n^{(k+2)}(x) &= \lambda_k(x) f_n^{'}(x) + s_k(x) f_n(x) \end{aligned} \tag{3}$$

Where

$$\lambda_k(x) = \lambda'_{k-1}(x) + s_{k-1}(x) + \lambda_0(x)\lambda_{k-1}(x),$$
$$s_k(x) = s'_{k-1}(x) + s_0(x)\lambda_{k-1}(x) \qquad (4)$$

Which are called the recurrence relation of (2). Taking the ratio of the $(k+2)th$ and $(k+1)th$ derivatives we get

$$\frac{d}{dx}\ln[f_n^{(k+1)}(x)] = \frac{f_n^{(k+2)}(x)}{f_n^{(k+1)}(x)} = \frac{\lambda_k(x)[f'_n(x) + \frac{s_k(x)}{\lambda_k(x)}f_n(x)]}{\lambda_{k-1}(x)[f'_n(x) + \frac{s_{k-1}(x)}{\lambda_{k-1}(x)}f_n(x)]} \qquad (5)$$

Assuming that for sufficiently large $k$

$$\frac{s_k(x)}{\lambda_k(x)} = \frac{s_{k-1}(x)}{\lambda_{k-1}(x)} = \alpha(x) \qquad (6)$$

holds, which is the "asymptotic aspect" of the method, equation (5) reduces to

$$\frac{d}{dx}\ln[f_n^{(k+1)}(x)] = \frac{f_n^{(k+2)}(x)}{f_n^{(k+1)}(x)} = \frac{\lambda_k(x)}{\lambda_{k-1}(x)} \qquad (7)$$

Substituting $\lambda_k(x)$ from equation (4) and then using $\alpha(x)$ in the right hand side of equation (7) we obtain

$$f^{(k+1)}(x) = C\ \exp\left(\int \frac{\lambda_k(x)}{\lambda_{k-1}(x)}dx\right) = C_1\lambda_{k-1}(x)\exp\left(\int^x [\alpha(x_1) + \lambda_0(x_1)]dx_1\right) \qquad (8)$$

in which $C_1$ is the integration constant. Inserting equation (8) into equation (3) and solving for $f(x)$, we obtain the general solution of equation (2) as

$$f_n(x) = \exp(-\int^x \alpha(x)dx_1)[C_2 + C_1\int^x \exp(\int^{x_1}[\lambda_0(x_2) + 2\alpha(x_2)]dx_2)dx_1] \qquad (9)$$

The energy eigenvalues can be determined by the quantization condition given by the termination condition in equation (6). Thus, one can write the quantization condition combined with Equation (4) as

$$\Delta_k(x) = \lambda_k(x) s_{k-1}(x) - \lambda_{k-1}(x) s_k(x) \tag{10}$$

If the eigenvalue problem is an exactly solvable problem, an explicit expression for energy can directly be obtained from the roots of this equation that depends only on the eigenvalues $E$. In this case, for any given quantum number $n$, the energy eigenvalues can be calculated from the roots of equation (10) at some suitable $x_0$ point. The chosen value of $x_0$ is arbitrary in principle and can be critical only to the speed of the convergence of the method. This starting value may be determined generally as the minimum value of the potential or the maximum value of the asymptotic wave function [19].

## SOLUTION AND ENERGY SPECTRUM

For a given energy $E$ and angular momentum $l$, time-independent radial Schrödinger equation for the reduced radial wave function, $R(r)$, can be written as

$$\left( \frac{-1}{2} \frac{d^2 R(r)}{dr^2} + \frac{l(l+1)}{2r^2} R(r) \right) + V(r) R(r) = E R(r) \tag{11}$$

Where the potential $V(r)$ given in equation (1); and we have adopted the atomic units $h = m = 1$.

In order to solve equation (11) for $l \neq 0$, we need to apply the approximation scheme to the centrifugal term given by [21]

$$\frac{1}{r^2} \approx \lambda^2 \left( \frac{1}{12} + \frac{e^{\lambda r}}{(e^{\lambda r} - 1)^2} \right) \tag{12}$$

Substituting equation (12) into equation (11), we obtain the following equation

$$\left( \frac{-1}{2} \frac{d^2 R(r)}{dr^2} + \lambda^2 \frac{l(l+1)}{2} \left( \frac{1}{12} + \frac{e^{\lambda r}}{(e^{\lambda r} - 1)^2} \right) R(r) \right) + V(r) R(r) = E R(r) \quad (13)$$

Now introducing the change in variable $x = 1 - 2 e^{-\lambda r}$ whose range is between −1 and +1, transforms equation (13) into the desired Equation (1) where we can apply the AIM with

$$\lambda_0(x) = \frac{1}{1-x} \quad (14)$$

$$s_0(x) = \frac{1}{12 \lambda^2 (x+1)^2 (x-1)^2} \left( 12 V_0 x^3 + \eta x^2 + \xi x + \nu \right) \quad (15)$$

Where

$$\eta = l(l+1) \lambda^2 + 12 (2\gamma - 1) V_0 - 24 E \quad (16)$$

$$\xi = -22 l(l+1) \lambda^2 - 12 V_0 - 48 E \quad (17)$$

And $\quad \nu = 25 l(l+1) \lambda^2 + (12 - 24 \gamma) V_0 - 24 E \quad (18)$

In order to obtain the energy eigenvalues from equation (13), using equation (4), we obtain the $\lambda_k(x)$, and $s_k(x)$ in terms of $\lambda_0(x)$, and $s_0(x)$. Then, using the quantization condition of the method given by equation (10), we obtain the energy eigenvalues. This straightforward application of AIM gives us the energy eigenvalues. We have observed that the energy eigenvalues converge within a reasonable number of iteration. This result agrees with the principle of AIM; as the number of iteration increases, the method should converge and should not oscillate.

## RESULTS AND DISCUSSION

The arbitrary $\ell$ - state solutions of the Schrödinger equation with short rang three parameter central potential have been obtained. To show the accuracy of our results, we have calculated the eigenvalues numerically for many $\ell$ - states and found that the results obtained by equation (10) are in good agreement with those obtained by other methods.

Our numerical results for the short rang three parameters central potential are listed in Tables 1, 2, 3. They show a comparison between the CSM and AIM eigen states. The results are in full agreement in the ground states, and agree very well for the higher states.

We used the data in table 4 to draw the relation between the angular momentum $\ell$ and the corresponding eigen states. Figure (1) shows that the ground states $E_o$ and $E_1$ are very deeply bounded. Figure (2) shows the same behavior for higher eigen states. As the angular momentum increases the energy eigen states increased until $\ell$ is greater than 2, we reached like a continuous state or we can say that the potential is not affecting any more.

## CONCLUSION

We used the AIM as a powerful engine to solve the Schrödinger equation for a short range three parameters potential with any angular momentum values. Our results are considered to be excellent compared to the CSM. We also introduced a new idea by studying the change of angular momentum with energy eigenvalues and clarifying the relation by Figure (1) and Figure (2). This relation was not being able to reach by other methods especially for higher $\ell$ values.

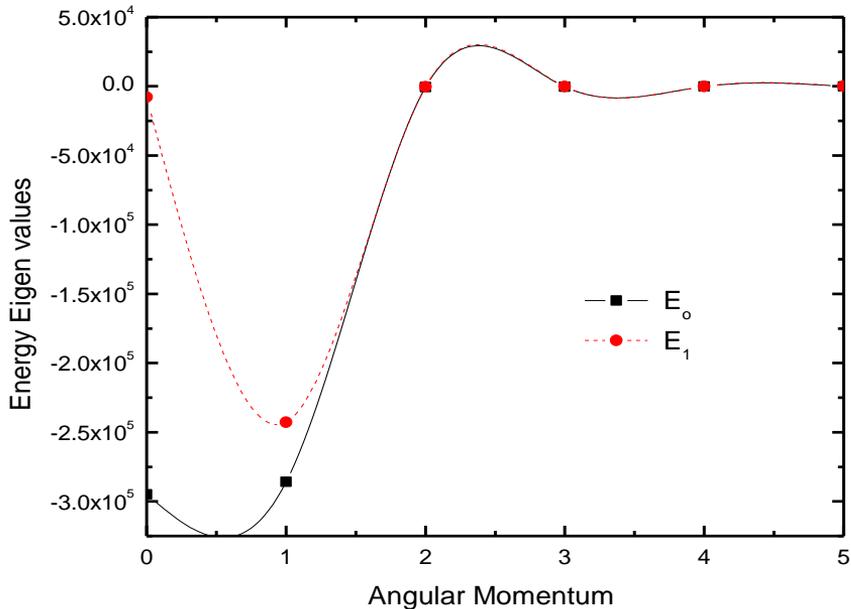

Figure 1: the energy eigenvalues as function of angular momentum $\ell$ for the gound ($E_o$) and first ($E_1$) excited states.

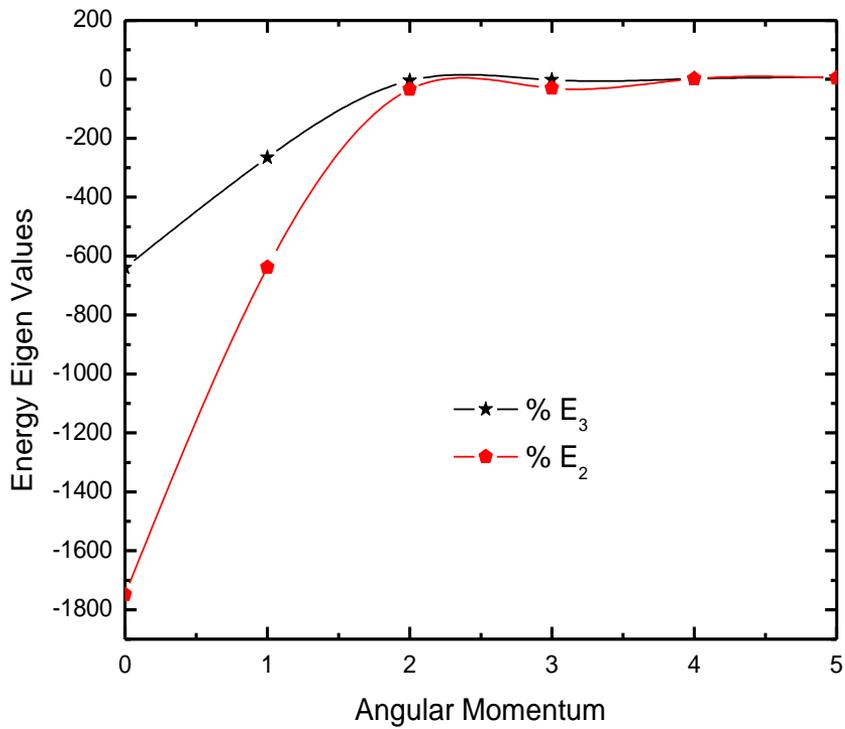

Figure 2: the energy eigenvalues as function of angular momentum $\ell$ for the second ($E_2$) and third ($E_3$) excited states.

**Table 1**: A comparison of the energy eigenvalues $E_n$ of the potential (1) obtained here by the AIM and compared to those obtained by the complex scaling Method (CSM) in [17]. We took $V_0 = -100, \gamma = 3/10, \lambda = \sqrt{2}$ and for various values of the angular momentum $\ell$.

| $\ell$ | n | $E_n$ by CSM | $E_n$ by AIM |
|---|---|---|---|
| 0 | 0 | −1094.42109160 | -1094.42109160 |
|   | 1 | −187.97359168 | -187.973591677 |
|   | 2 | −36.02622806 | -36.0262280564 |
| 1 | 0 | −185.38841241 | -185.3881967 |
|   | 1 | 33.95322592 | -33.95168758 |
| 2 | 0 | −29.64332195 | -29.63997992 |

**Table 2**: A comparison of the energy eigenvalues $E_n$ of the potential (1) obtained here by the AIM and compared to those obtained by the complex scaling Method (CSM) in [17]. We took $V_0 = -160, \gamma = 5/10, \lambda = \sqrt{2}$ and for various values of the angular momentum $\ell$.

| $\ell$ | n | $E_n$ by CSM | $E_n$ by AIM |
|---|---|---|---|
| 0 | 0 | −1406.11040577 | -1406.11040577 |
|   | 1 | −223.29635015 | -223.296350148 |
|   | 2 | −27.18320883 | -27.1832088304 |
| 1 | 0 | −219.66959141 | -219.6694254 |
|   | 1 | −24.21918006 | -24.21795546 |
| 2 | 0 | −18.01714564 | -18.01447261 |

**Table 3:** A comparison of the energy eigenvalues $E_n$ of the potential in equation (1) obtained here by the AIM and compared to those obtained by the complex scaling Method (CSM) in [17]. We took $V_0 = -200, \gamma = 7/10, \lambda = \sqrt{2}$ for various values of the angular momentum $\ell$.

| $\ell$ | n | $E_n$ by CSM | $E_n$ by AIM |
|---|---|---|---|
| 0 | 0 | −679.95986643 | -679.959866430 |
|   | 1 | −32.96147955 | -32.961479552 |
| 1 | 0 | −27.29186980 | -27.29153038 |

**Table 4:** The energy eigenvalues for different angular momentum $\ell$; calculated by AIM. The values of the potential parameters are $V_0 = -200, \gamma = 7/10, \lambda = \sqrt{2}$

| $\ell$ | $E_0$ | $E_1$ | $E_2$ | $E_3$ |
|---|---|---|---|---|
| 0 | -2.948040331×10$^5$ | -7813.547139 | -1748.849560 | -640.0248682 |
| 1 | -2.859760644×10$^5$ | -2.428306994×10$^5$ | -637.9431013 | -265.3244481 |
| 2 | -633.7598994 | -261.5672734 | -33.72115660 | -4.266342233 |
| 3 | -255.8552832 | -99.68104017 | -29.72230919 | -1.541952406 |
| 4 | -92.89079567 | -24.18745837 | 2.003775523 | 2.116254874 |
| 5 | -16.85605236 | 2.897862067 | 4.531566676 | 6.360648382 |